# Elasticity, Flexibility and Ideal Strength of Borophenes

Zhuhua Zhang, Yang Yang, Evgeni S. Penev and Boris I. Yakobson*

*Department of Materials Science and NanoEngineering, and Department of Chemistry, Rice University, Houston, Texas 77005, United States*
\*email: biy@rice.edu

**ABSTRACT**: We study the mechanical properties of two-dimensional (2D) boron—borophenes—by first-principles calculations. The recently synthesized borophene with 1/6 concentration of hollow hexagons (HH) is shown to have in-plane modulus $C$ up to 210 N/m and bending stiffness as low as $D = 0.39$ eV. Thus, its Foppl–von Karman number per unit area, defined as $C/D$, reaches 568 nm$^{-2}$, over twofold higher than graphene's value, establishing the borophene as one of the most flexible materials. Yet, the borophene has a specific modulus of 346 m$^2$/s$^2$ and ideal strengths of 16 N/m, rivaling those (453 m$^2$/s$^2$ and 34 N/m) of graphene. In particular, its structural fluxionality enabled by delocalized multi-center chemical bonding favors structural phase transitions under tension, which result in exceptionally small breaking strains yet highly ductile breaking behavior. These mechanical properties can be further tailored by varying the HH concentration, and the boron sheet without HHs can even be stiffer than graphene against tension. The record high flexibility combined with excellent elasticity in boron sheets can be utilized for designing composites and flexible systems.

Positioned between beryllium and carbon in the periodic table, boron is a key element that has chemical features of both metal and non-metal. Because of this unique nature, a rich variety of bonding configurations can form between boron atoms, ranging from normal two-center two-electron bonds to up to seven-center two-electron bonds,[1] enabling boron to form a large number of allotropes and compounds with most elements. Since the advent of graphene,[2] two-dimensional (2D) materials that are one or several atoms thick reign the current field of materials research. As boron has demonstrated striking similarity to carbon, forming planar clusters[1, 3-8], cage-like fullerences[9-15] and 1D nanotubes [7, 16-21], extensive theoretical efforts have been devoted to exploring graphene analogues of boron—borophenes[22-26]. Unlike graphene or hexagonal boron nitride (h-BN) that have exclusively stable honeycomb lattice, the borophene is predicted to be polymorphic[27] with numerous states near the ground-state energy line, due to a highly variable network of hollow hexagons (HHs) in a reference triangular lattice. Towards the synthesis of borophenes, several theoretical works[28-29] have suggested using metal substrates, such as Ag or Cu that may screen only a few of 2D boron structures. Very recently, two independent experiments have reported the successful synthesis of borophenes on Ag(111) by using molecular beam epitaxy methods [30-31].

Because of reduced dimensionality, borophenes show qualitatively different properties from bulk boron. For example, while all existing 3D boron allotropes are semiconducting at standard conditions, borophenes show metallic characteristic[30-32], which also distinguishes them from other well studied 2D materials usually having bandgaps. Among the properties borophenes may offer, their mechanical properties are of particular interest and importance. First, since boron is lighter than most elements, borophenes have lower mass density than other 2D materials. This opens an intriguing possibility of using borophenes as reinforcing elements for designing composites, provided that their in-plane stiffness and ideal strength are satisfactorily high. Second, 2D materials exhibit high levels of flexibility against out-of-plane deformation and are suitable for fabricating flexible devices. The metallic borophenes are specially potential for making flexible electrode and contact, both being crucial for nanoelectronics. Fully materializing the borophenes' potential into these applications is contingent upon knowing their elastic properties, ideal strength and structural flexibility. Unlike graphene and h-BN sheet made of two-electron two-center bonding, borophenes are braided by delocalized multicenter two-electron bonding[33]. Little from extensive study of mechanics in graphene and h-BN sheet can be generalized to borophenes, where qualitatively new phenomena could be anticipated.

Here, we show by first-principles calculations that the experimentally realized borophene, with a HH concentration of v=1/6 (defined as v=$m/N$, where $m$ is the number of HHs in a unit cell of $N$ triangular lattice sites), is a type of material that combines supreme flexibility and excellent in-plane elasticity. The borophene has an out-of-plane bending stiffness fourfold lower than graphene but its elastic moduli exceeds one-half of graphene's value. As a result, borophene possesses a record high Foppl–von Karman number per unit area[34-35] as 568 nm$^{-2}$, establishing it as one of the most flexible materials. Despite the extreme flexibility, the ideal strength of borophene can be over 16 N/m, higher than those of the best known polymer materials, and its specific modulus is up to 346 m$^2$/s$^2$, rivaling those of graphene. In contrast to other 2D materials, the borophene can relieve tensile stress by increasing HH concentration, manifested as a series of tension-induced phase transitions, and thereby withstand a tensile load ~10 N/m at strains as high as 40%. Moreover, the flexibility, elasticity and strength of borophene can be tailored by varying

HH concentration, which may be controlled using different metal substrates in synthesis. These mechanical properties beyond those of existing 2D materials will promote the applications of borophenes in nanoelectromechanical systems and flexible electronics.

All the calculations are implemented in Vienna *Ab-initio* Simulation Package (VASP) code.[36-37] We employ ultrasoft pseudo-potentials for the core region and spin-unpolarized density functional theory (DFT) based on the generalized gradient approximation of Perdew–Burke–Ernzerhof [38] functional. The kinetic energy cutoff of the plane-wave expansion is set to be 400 eV. The borophenes are simulated using supercells with a periodic boundary condition. The vacuum region between two borophene layers in adjacent periodic images is fixed to 15 Å in order to eliminate spurious interaction. For evaluating the bending stiffness, infinite boron nanotubes with different diameter are simulated with a primitive tube cell. The Brillouin-zone is densely sampled for integration, with approximately the same $k$-point density among different-sized supercells. Uniaxial tensile strain is applied along x and y directions (see definition later) of the simulated box. The borophenes are stretched step by step, and, meanwhile, the dimension is allowed to shrink in the orthogonal direction. To allow possible reconstructions during tension, we use 1×2 and 3×2 supercells for the $v_{1/6}$ and triangular sheets, respectively; a part of boron atoms are displaced along the stretching direction from their ideal position prior to structural relaxation. Different levels of atomic displacement are considered in order to select the optimal structure at each applied strain $\varepsilon$. This method has proven reliable for evaluating the strength of single-walled carbon nanotubes.[39] All the atomic positions are fully relaxed using the conjugate-gradient method until the force on each atom is less than 0.01 eV/Å. The energy barriers for bond rotation are calculated using the climbing image-nudged elastic-band method.[40]

The experimentally realized borophenes were characterized to be in different lattice phases. The triangular (v=0) and $v_{1/6}$ sheets were proposed as atomic models for a stripped phase[30, 41] while the $v_{1/5}$ sheet was proposed for a phase showing a brick-wall pattern[31] in scanning tunneling microscopy images. The relaxed triangular sheet displays a washboard-like buckling structure due to an excess of electrons (Figure 1a), whereas the $v_{1/6}$ and $v_{1/5}$ sheets are purely planar because of their high HH concentrations making them both electron-deficient (Figure 1b and 1c). In addition, we also consider the planar α (i.e. v=1/9) and $v_{1/8}$ sheets, two of the most stable structures in vacuum[27], and a buckled $v_{1/12}$ sheet (Figure S1), one of the most stable structures on Au(111) [29]. For ease of discussion, we define the lattice orientation along straight boron chains as x direction and that perpendicular to the chains as y direction (see Figure 1).

In general, engineering strain can be applied in an arbitrary lattice orientation; to examine possible anisotropy, we consider two basic cases, that is, the strains applied along the x and y directions. The calculated stress-strain curves of all borophenes are linearly elastic at small strains. The elastic response can be either isotropic or anisotropic depending on structural symmetry. The Young's modulus, $C$, and Poisson ratio, $\sigma$, of different boron sheets are summarized in Table I, together with those of other typical 2D materials for comparison. For the $v_{1/6}$ sheet, $C$ are 189 N/m and 210 N/m along x and y directions, respectively, lower than 342 N/m for graphene [42]; the corresponding $\sigma$ are 0.15 and 0.17, respectively, almost identical to 0.17 for graphene. Interestingly, $C$ of planar B sheet is quite insensitive to v, varying from 189 N/m to 216 N/m in the x direction and from 208 N/m to 222 N/m in the y direction upon changing v from 1/5 to 1/9. This insensitivity can be understood by taking borophenes as a binary system composed of HHs and filled hexagons. With applied in-plane strain, the bond deformation around the HHs is more pronounced than that in the filled hexagons, as shown in Figure 2 for the $v_{1/6}$ and α sheets, suggesting that the HHs are mechanically "softer" than the filled hexagons. Since the applied strain is highly concentrated around the HHs, they dominates the elasticity of entire sheets and result in the insensitivity of $C$ and $\sigma$ to v. In the triangular sheet with no HH, the bond deformation is uniform; $C$ becomes outstandingly high along the ridges, but decreases to 163 N/m across the ridges because the tensile deformation is largely contributed by bond rotation in this direction.

Since the HHs tend to result in a planar sheet and govern its elasticity, the 2D B sheets with HHs display insignificant anisotropy. The α sheet is even isotropic due to its high symmetry. Likewise, the Poisson ratio is almost isotropic upon varying v from 1/5 to 1/9. Interestingly, the triangular sheet has zero Poisson ratio across the ridges and even negative Poisson ratio when stretching along the ridges. The negative Poisson ratio results from the Poisson effect along the out-of-plane direction, that is the applied strain reduces the buckling amplitude of the sheet; in turn, instead of shrinking, the ridge-to-ridge distance is increased markedly due to a much lower in-plane modulus across the ridges than along the ridges. The negative Poisson ratio should also exist in other buckled borophenes with low HH concentrations. Earlier theories have reported negative Poisson ratio in phosphorene[43] and a newly predicted $Be_5C_2$ monolayer[43-44], both with a large out-of-plane buckling.

Among the known 2D materials, borophene should be the lightest one since only a few elements can be simpler than boron. This motivates us to calculate the specific modulus, i.e. the elastic modulus per mass density $C/\rho$. Figure 3a summarizes the specific modulus of borophenes. For the $v_{1/6}$ sheet, the specific modulus is 346 $m^2/s^2$ in the y direction, which reaches 76% of graphene's value and 95% of h-BN's value. Meanwhile, this value is 4-20 times higher than those of other 2D materials and twofold higher than boron fibers[45], opening the possibility of using borophenes as reinforcing elements. It is of practical importance that the specific modulus of borophene does not depend much on v, nor on lattice orientation. An exception again arises at the triangular sheet, which has a record high value of 516 $m^2/s^2$ along the ridges but a much smaller value of 211 $m^2/s^2$ across the ridges.

Having revealed the in-plane elasticity of borophenes, we proceed to examine their out-of-plane bending stiffness $D$, which is determined by fitting the calculated bending energy per unit area $E_{ben}$ of a boron nanotube as a function of tube radius $r$, based on an analytical expression $E_{ben} = Dr^{-2}/2$. In this manner, we obtain $D$ = 1.42 eV for graphene, in good agreement with previous DFT-level values[46-50]. For the 2D B polymorphs, $D$ turns out to rather small in the x and y directions, as listed in Table I. In particular, the $v_{1/6}$ sheet has $D$ as low as 0.39 eV along the HH rows, smaller than the reported value of any 2D material. The reason for this record small $D$ is twofold. First, the HHs are aligned into parallel rows, which facilitate bending along the rows due to decreased bond density therein. Second, the row-to-row spacing in the $v_{1/6}$ sheet is unique in a manner that optimizes the length of horizontal B-B bonds in the HH rows and, meanwhile, minimizes the energy cost of bending more rigid filled hexagons. This point becomes clearer when comparing structures of the $v_{1/5}$ and $v_{1/12}$ sheets, both with HH rows too. In the $v_{1/5}$ sheet, the B-B bonds in the HH rows are shorter by 0.03 Å and hence stronger than those in the $v_{1/6}$ sheet; its $D$ thus rises to 0.56 eV. Yet, in the $v_{1/12}$ sheet, the HH rows are sparsely spaced and the wide segments made of filled hexagons are more involved in bending deformation (Figure S1), which results in an even higher $D$ of 0.92 eV.

Generally, decreasing v close to 0 leads to an increase of $D$ and results in more evident anisotropy, since the corresponding borophenes are increasingly buckled. The triangular sheet has a maximum buckling of 0.87 Å, which increases $D$ to 1.39 eV along the ridges and up to 4.92 eV across the ridges. This bending stiffness is comparable to other 2D materials with finite thickness, such as 5.21 eV for phosphorene and 9.14 eV for single-layer $MoS_2$ (close to 9 eV in an earlier work[51]).

The combined excellent in-plane stiffness and exceptionally small flexural rigidity render borophenes as an atomic membrane that is easy to bend yet hard to stretch. The material parameter characterizing this behavior is the Foppl–von Karman number per unit area, $\gamma$, defined as $C/D$. Compared to specific modulus, $\gamma$ varies much more widely from material to material (Figure 3b). The $v_{1/6}$ sheet has a highest $\gamma$ of 568 $nm^{-2}$, which is over two times higher than 234 $nm^{-2}$ for graphene and forty times higher than 14 $nm^{-2}$ for the $MoS_2$ monolayer. The $v_{1/6}$ sheet thus represents one of the most flexible materials, yet with considerably high in-plane modulus. Varying $v$ can significantly modulate $\gamma$, which decreases to 377 $nm^{-2}$ for the $v_{1/5}$ sheet and 34 $nm^{-2}$ for the triangular sheet along the ridges. It is worth mentioning that the $v_{1/5}$, $v_{1/8}$, and $\alpha$ sheets still have higher $\gamma$ than graphene, independent of lattice orientation.

The unprecedented flexibility of borophene can be further visualized by wrapping it around an object and then examining its morphology. For demonstration, we use a BN nanotube, around which the $MoS_2$, graphene and $v_{1/6}$ sheet are wrapped, respectively. The 2D materials will bend either locally or uniformly, depending on their bending stiffness $D$ and curvature $r^{-1}$. The energy change per unit area of a 2D material wrapping around a BN nanotube can be expressed as $\Delta E = E_{bend} + E_{vdw}$, where $E_{bend} = Dr^{-2}/2$ is bending energy and $E_{vdw}$ is van der Waals contribution. When $E_{vdw} > E_{bend}$, the 2D material tends to bend uniformly around the nanotube. Note that $E_{vdw}$ is a constant but $E_{bend}$ increases with decreasing tube radius. Therefore, there is a critical tube radius, below which the 2D material starts to bend locally. By calculating $E_{vdw}$, the critical tube radius is determined to be 1.8 nm for $MoS_2$ monolayer and 0.9 nm for graphene but drops to 0.3 nm for the $v_{1/6}$ sheet. Relaxed structures of $MoS_2$, graphene and $v_{1/6}$ sheet wrapping around a (7,0) nanotube ($r$ = 0.3 nm) are shown in Figure S2. In accord with our analysis, both the $MoS_2$ and graphene bend locally, but the $v_{1/6}$ sheet is fully wrapped around the nanotube. The demonstrated extreme flexibility would allow the $v_{1/6}$ sheet to precisely follow a solid surface with nanoscale roughness, to an extent that the interspaces between the sheet and surface are minimized to impede intercalation of external chemical species. This feature may invite potential applications, such as *coating* for anticorrosion if a chemically inert capping layer is additionally deposited.

Another important mechanical parameter is the ideal strength of borophenes. Unlike other 2D materials with covalent bonds, borophenes are made of multi-center bonds, which may bring out distinctive mechanical response in the elastic limit. The ideal tensile stress versus engineering strain for the triangular sheet is shown in Figure 4a. At small strains, the sheet exhibits linear stress-strain relationship, with notable elastic anisotropy. As the strain increases, the stress-strain behaviors become increasingly nonlinear and show enhanced anisotropy. The peak stress reaches 20.9 N/m along the ridges at $\varepsilon_{xx}$=8.7% and 12.2 N/m at $\varepsilon_{yy}$=14.3% across the ridges. The peak strengths and critical strains of the triangular sheet are remarkably close to previously reported values attained by examining the phonon instability[52], confirming the validity of our method based on atomic displacement in predicting the ideal strength. Since other 2D B sheets have intrinsic negative phonon frequencies that may obscure the appearance of negative frequency induced by strain, we continue to use our method to examine their ideal strength. Figure 4b presents the results for the $v_{1/6}$ sheet, whose peak strength is 16.4 N/m at $\varepsilon_{xx}$=12.5% across the HH rows and 15.4 N/m at $\varepsilon_{yy}$=10.6% along the HH rows, showing a little anisoropy. As long as the HHs are included, the ideal strength of borophene changes limitedly upon varying v, as evidenced by the similar values in the α and $v_{1/8}$ sheets (Figure 3c). An implication of this result is that the strength of borophenes can be insensitive to point defects, such as monovacancy and adatoms (equivalent to changing HHs), in contrast to other 2D materials where these defects markedly degrade their mechanical performance. While the ideal strengths of borophenes are inferior to graphene and h-BN sheet[53], they are significantly higher than other 2D materials, such as $MoS_2$[54] and phosphorene[55] (Figure 3c). Of more interest is that all the strengths of borophenes are reached at exceptionally small critical strains (8%~15%), compared to those of other 2D materials (Figure 3d). The small critical strains are attributed to an unusual response of the multi-center bonding at elastic limit, as discussed below.

Materials at critical tensile strain are usually ensued by structural failure. However, this is not the case for borophenes, which undergo structural phase transitions under tension. Figure 4b shows that the $v_{1/6}$ sheet stretched across the HH rows is transformed into a new $v_{1/7}$ sheet when $\varepsilon_{xx}$ > 12.5% and further into an B monolayer comprised of octagons, squares and triangles as $\varepsilon_{xx}$ is increased to 32% (Figure 4b, top inserts). When stretched along the HH rows, the structural distortion starts to appear in the rows of filled hexagons at $\varepsilon_{yy}$=10.6%; then the structure is transformed to a new B sheet made of triangles and octagons as $\varepsilon_{yy}$ is increased to 30% (Figure 5b, bottom inserts). The phase transition endows borophenes with high toughness against tensile loading. For example, the $v_{1/6}$ sheet can resist a load of 11 N/m and 6 N/m at a tensile strain as high as 36% applied across and along the HH rows, respectively, at which even the strongest graphene has been broken in our simulations. The tension-induced phase transition is a common behavior in the boron sheets with HHs. Our calculations show that the $v_{1/8}$ and α sheets can all be transformed into new sheets with lower HH concentrations (Figure S3), giving them a large loading capability even under extremely high strain (>30%). We have verified that these key results can be reproduced in simulations using a larger supercell (Figure S4).

The strain-induced phase transition is rare in 2D materials, and it benefits from the variable B coordination from 3 to 6 as well as structural fluxionality of borophenes enabled by delocalized multicenter two-electron B-B bonds. The phase transition involves atomic rearrangements via bond rotations. We thus searched for the transition states of bond rotations in the $v_{1/6}$ sheet using nudged elastic-band method. The energy barrier is 2.28 eV for rotating a bond in the HH row (marked by black thick line) and 1.7 eV for rotating a bond (marked by blue thick line) tilted with respect to the HH row (Figure 4c). These barriers are over three-fold lower than that (~9 eV) for Stone-Wales bond rotation in graphene, indicating that the bonds are much easier to flip in borophenes. Of more significance is that the two barriers can drop to almost zero at $\varepsilon$=10% applied along and across the HH rows, respectively (Figure 4d). Accordingly, the reaction energies, i.e. the energy difference between the initial and final states, become highly negative at $\varepsilon$=10% to help drive the bond flip. We have also noted that in some planar boron clusters the inner B atoms can rotate against the periphery of the clusters with almost no barrier.[56] Another point worth mentioning is that pentagons and heptagons, which are normal products of a Stone-Wales bond rotation in carbon materials, relax into triangles or fused hexagons in borophene. In this manner, the stress created by bond rotation is relieved in the final state, whose energy is thereby greatly reduced.

In contrast, the triangular sheet shows a brittle fracture at a strain of ~14%, above which the stress sharply drops to zero (Figure 5a). The brittle fracture can be understood from two aspects: i) Due to excessive electrons, the triangular sheet has out-of-plane buckling, which is decreased by applied tensile strain; the decreased buckling diminishes the mixing of in-plane and out-of-plane orbitals and thereby drives more of the excessive electrons to occupy the in-plane antibonding states, ending up with severely weakened B-B bonds. ii) The dangling bonds along the cleaved edges are self-passivated due to the variability of B coordination; the self-passivation is particularly efficient at the cleaved flat B edge (insets in Figure 5a), which naturally exists in many planar B molecules [1, 4, 17].

In conclusion, we have performed comprehensive first-principles analyses of the mechanical properties of borophenes, which are shown to combine excellent elasticity, unprecedented flexibility and high ideal strength. In particular, the borophene with a HH concentration of 1/6 has a bending stiffness fourfold lower than the value of graphene, attributed to optimally spaced HH rows. Yet, the Young's modulus of the $v_{1/6}$ sheet reaches up to 210 N/m, over 60% of the graphene's value. Thus, the $v_{1/6}$ sheet possesses the highest Foppl–von Karman number per unit area, i.e. the ratio between in-plane modulus and bending stiffness, featuring it as a material that most easily bends and crumples than it stretches. The high in-plane elasticity of the borophene is further aided by its low mass density, leading its specific modulus of 346 $m^2/s^2$ to be close to that of graphene. Of more surprise is that this flexible material has an ideal strength of 16 N/m, only secondary to graphene and h-BN sheet but being several times higher than $MoS_2$ monolayer, phosphorene and silicene. Being made of delocalized multi-center bondings, the borophene does not fracture after the peak strength but experiences strain-induced structural phase transitions that toughen the materials further, to an extent that it still can resist the same levels of loading as original even at a strain over 35%. All these mechanical properties of the borophene can be further adjusted by varying the HH concentration. Our findings offer new insight into the mechanical response of multi-center two-electron bonds and suggest potential applications of borophenes in making composites and flexible devices.

**References**

1. Sergeeva, A. P.; Popov, I. A.; Piazza, Z. A.; Li, W.-L.; Romanescu, C.; Wang, L.-S.; Boldyrev, A. I., Understanding boron through size-selected clusters: structure, chemical bonding, and fluxionality. *Accounts of chemical research* **2014,** *47* (4), 1349-1358.
2. Novoselov, K. S.; Geim, A. K.; Morozov, S. V.; Jiang, D.; Zhang, Y.; Dubonos, S. V.; Grigorieva, I. V.; Firsov, A. A., Electric field effect in atomically thin carbon films. *Science* **2004,** *306* (5696), 666-669.
3. Li, W.-L.; Chen, Q.; Tian, W.-J.; Bai, H.; Zhao, Y.-F.; Hu, H.-S.; Li, J.; Zhai, H.-J.; Li, S.-D.; Wang, L.-S., The $B_{35}$ cluster with a double-hexagonal vacancy: a new and more flexible structural motif for borophene. *Journal of the American Chemical Society* **2014,** *136* (35), 12257-12260.
4. Piazza, Z. A.; Hu, H.-S.; Li, W.-L.; Zhao, Y.-F.; Li, J.; Wang, L.-S., Planar hexagonal $B_{36}$ as a potential basis for extended single-atom layer boron sheets. *Nature communications* **2014,** *5*, 6.
5. Sergeeva, A. P.; Piazza, Z. A.; Romanescu, C.; Li, W.-L.; Boldyrev, A. I.; Wang, L.-S., $B_{22}^-$ and $B_{23}^-$: All-boron analogues of anthracene and phenanthrene. *Journal of the American Chemical Society* **2012,** *134* (43), 18065-18073.

6. Li, W.-L.; Romanescu, C.; Jian, T.; Wang, L.-S., Elongation of planar boron clusters by hydrogenation: Boron analogues of polyenes. *Journal of the American Chemical Society* **2012,** *134* (32), 13228-13231.
7. Oger, E.; Crawford, N. R.; Kelting, R.; Weis, P.; Kappes, M. M.; Ahlrichs, R., Boron cluster cations: transition from planar to cylindrical structures. *Angewandte Chemie International Edition* **2007,** *46* (44), 8503-8506.
8. Huang, W.; Sergeeva, A. P.; Zhai, H.-J.; Averkiev, B. B.; Wang, L.-S.; Boldyrev, A. I., A concentric planar doubly π-aromatic $B_{19}^-$ cluster. *Nature Chemistry* **2010,** *2* (3), 202-206.
9. Szwacki, N. G.; Sadrzadeh, A.; Yakobson, B. I., $B_{80}$ fullerene: an ab initio prediction of geometry, stability, and electronic structure. *Physical review letters* **2007,** *98* (16), 166804.
10. Zhai, H.-J.; Zhao, Y.-F.; Li, W.-L.; Chen, Q.; Bai, H.; Hu, H.-S.; Piazza, Z. A.; Tian, W.-J.; Lu, H.-G.; Wu, Y.-B., Observation of an all-boron fullerene. *Nature Chemistry* **2014,** *6* (8), 727-731.
11. Sadrzadeh, A.; Pupysheva, O. V.; Singh, A. K.; Yakobson, B. I., The boron buckyball and its precursors: an electronic structure study. *The Journal of Physical Chemistry A* **2008,** *112* (51), 13679-13683.
12. Li, H.; Shao, N.; Shang, B.; Yuan, L.-F.; Yang, J.; Zeng, X. C., Icosahedral $B_{12}$-containing core–shell structures of $B_{80}$. *Chemical Communications* **2010,** *46* (22), 3878-3880.
13. Wang, L.; Zhao, J.; Li, F.; Chen, Z., Boron fullerenes with 32–56 atoms: Irregular cage configurations and electronic properties. *Chemical Physics Letters* **2010,** *501* (1), 16-19.
14. Zhao, J.; Wang, L.; Li, F.; Chen, Z., $B_{80}$ and other medium-sized boron clusters: Core−shell structures, not hollow cages. *The Journal of Physical Chemistry A* **2010,** *114* (37), 9969-9972.
15. Lv, J.; Wang, Y.; Zhu, L.; Ma, Y., $B_{38}$: An all-boron fullerene analogue. *Nanoscale* **2014,** *6* (20), 11692-11696.
16. Ciuparu, D.; Klie, R. F.; Zhu, Y.; Pfefferle, L., Synthesis of pure boron single-wall nanotubes. *The Journal of Physical Chemistry B* **2004,** *108* (13), 3967-3969.
17. Kiran, B.; Bulusu, S.; Zhai, H.-J.; Yoo, S.; Zeng, X. C.; Wang, L.-S., Planar-to-tubular structural transition in boron clusters: $B_{20}$ as the embryo of single-walled boron nanotubes. *Proceedings of the National Academy of Sciences of the United States of America* **2005,** *102* (4), 961-964.
18. Singh, A. K.; Sadrzadeh, A.; Yakobson, B. I., Probing properties of boron α-tubes by ab initio calculations. *Nano letters* **2008,** *8* (5), 1314-1317.
19. Liu, F.; Shen, C.; Su, Z.; Ding, X.; Deng, S.; Chen, J.; Xu, N.; Gao, H., Metal-like single crystalline boron nanotubes: synthesis and in situ study on electric transport and field emission properties. *Journal of Materials Chemistry* **2010,** *20* (11), 2197-2205.
20. Tian, J.; Xu, Z.; Shen, C.; Liu, F.; Xu, N.; Gao, H.-J., One-dimensional boron nanostructures: Prediction, synthesis, characterizations, and applications. *Nanoscale* **2010,** *2* (8), 1375-1389.
21. Bezugly, V.; Kunstmann, J.; Grundkötter-Stock, B.; Frauenheim, T.; Niehaus, T.; Cuniberti, G., Highly conductive boron nanotubes: transport properties, work functions, and structural stabilities. *ACS Nano* **2011,** *5* (6), 4997-5005.
22. Tang, H.; Ismail-Beigi, S., Novel precursors for boron nanotubes: the competition of two-center and three-center bonding in boron sheets. *Physical review letters* **2007,** *99* (11), 115501.
23. Yang, X.; Ding, Y.; Ni, J., Ab initio prediction of stable boron sheets and boron nanotubes: structure, stability, and electronic properties. *Physical Review B* **2008,** *77* (4), 041402.

24. Wu, X.; Dai, J.; Zhao, Y.; Zhuo, Z.; Yang, J.; Zeng, X. C., Two-dimensional boron monolayer sheets. *ACS Nano* **2012,** *6* (8), 7443-7453.
25. Zhang, Z.; Penev, E. S.; Yakobson, B. I., Two-dimensional materials: Polyphony in B flat. *Nature Chemistry* **2016,** *8* (6), 525-527.
26. Lu, H.; Mu, Y.; Bai, H.; Chen, Q.; Li, S.-D., Binary nature of monolayer boron sheets from ab initio global searches. *The Journal of chemical physics* **2013,** *138* (2), 024701.
27. Penev, E. S.; Bhowmick, S.; Sadrzadeh, A.; Yakobson, B. I., Polymorphism of two-dimensional boron. *Nano letters* **2012,** *12* (5), 2441-2445.
28. Liu, Y.; Penev, E. S.; Yakobson, B. I., Probing the Synthesis of Two‐Dimensional Boron by First‐Principles Computations. *Angewandte Chemie International Edition* **2013,** *52* (11), 3156-3159.
29. Zhang, Z.; Yang, Y.; Gao, G.; Yakobson, B. I., Two-Dimensional Boron Monolayers Mediated by Metal Substrates. *Angewandte Chemie International Edition* **2015,** *54* (44), 13022-13026.
30. Mannix, A. J.; Zhou, X.-F.; Kiraly, B.; Wood, J. D.; Alducin, D.; Myers, B. D.; Liu, X.; Fisher, B. L.; Santiago, U.; Guest, J. R., Synthesis of borophenes: Anisotropic, two-dimensional boron polymorphs. *Science* **2015,** *350* (6267), 1513-1516.
31. Feng, B.; Zhang, J.; Zhong, Q.; Li, W.; Li, S.; Li, H.; Cheng, P.; Meng, S.; Chen, L.; Wu, K., Experimental realization of two-dimensional boron sheets. *Nature Chemistry* **2016,** *8* (6), 563-568.
32. Feng, B.; Zhang, J.; Liu, R.-Y.; Iimori, T.; Lian, C.; Li, H.; Chen, L.; Wu, K.; Meng, S.; Komori, F., Direct evidence of metallic bands in a monolayer boron sheet. *Physical Review B* **2016,** *94* (4), 041408.
33. Galeev, T. R.; Chen, Q.; Guo, J.-C.; Bai, H.; Miao, C.-Q.; Lu, H.-G.; Sergeeva, A. P.; Li, S.-D.; Boldyrev, A. I., Deciphering the mystery of hexagon holes in an all-boron graphene α-sheet. *Physical Chemistry Chemical Physics* **2011,** *13* (24), 11575-11578.
34. Foppl, A., *Vorlesungen uber technische Mechanik*. B. G. Teubner: 1905.
35. von Karman, T., *Festigkeitsproblem im Maschinenbau*. Encyklopadie der Mathematischen Wissenschaften: 1910; Vol. 4.
36. Kresse, G.; Furthmüller, J., Efficiency of ab-initio total energy calculations for metals and semiconductors using a plane-wave basis set. *Computational Materials Science* **1996,** *6* (1), 15-50.
37. Kresse, G.; Furthmüller, J., Efficient iterative schemes for ab initio total-energy calculations using a plane-wave basis set. *Physical Review B* **1996,** *54* (16), 11169.
38. Perdew, J. P.; Burke, K.; Ernzerhof, M., Generalized gradient approximation made simple. *Physical review letters* **1996,** *77* (18), 3865.
39. Dumitrica, T.; Hua, M.; Yakobson, B. I., Symmetry-, time-, and temperature-dependent strength of carbon nanotubes. *Proceedings of the National Academy of Sciences* **2006,** *103* (16), 6105-6109.
40. Henkelman, G.; Uberuaga, B. P.; Jónsson, H., A climbing image nudged elastic band method for finding saddle points and minimum energy paths. *The Journal of chemical physics* **2000,** *113* (22), 9901-9904.
41. Dewhurst, R. D.; Claessen, R.; Braunschweig, H., Two‐Dimensional, but not Flat: An All‐Boron Graphene with a Corrugated Structure. *Angewandte Chemie International Edition* **2016,** *55* (16), 4866-4868.

42. Lee, C.; Wei, X.; Kysar, J. W.; Hone, J., Measurement of the elastic properties and intrinsic strength of monolayer graphene. *Science* **2008,** *321* (5887), 385-388.
43. Wang, Y.; Li, F.; Li, Y.; Chen, Z., Semi-metallic $Be_5C_2$ monolayer global minimum with quasi-planar pentacoordinate carbons and negative Poisson's ratio. *Nature communications* **2016,** *7*.
44. Jiang, J.-W.; Park, H. S., Negative poisson's ratio in single-layer black phosphorus. *Nature communications* **2014,** *5*.
45. Riley, M.; Whitney, J., Elastic properties of fiber reinforced composite materials. *Aiaa Journal* **1966,** *4* (9), 1537-1542.
46. Kudin, K. N.; Scuseria, G. E.; Yakobson, B. I., $C_2F$, BN, and C nanoshell elasticity from ab initio computations. *Physical Review B* **2001,** *64* (23), 235406.
47. Muñoz, E.; Singh, A. K.; Ribas, M. A.; Penev, E. S.; Yakobson, B. I., The ultimate diamond slab: GraphAne versus graphEne. *Diamond and Related Materials* **2010,** *19* (5), 368-373.
48. Ma, T.; Li, B.; Chang, T., Chirality-and curvature-dependent bending stiffness of single layer graphene. *Applied Physics Letters* **2011,** *99* (20), 201901.
49. Koskinen, P.; Kit, O. O., Approximate modeling of spherical membranes. *Physical Review B* **2010,** *82* (23), 235420.
50. Wei, Y.; Wang, B.; Wu, J.; Yang, R.; Dunn, M. L., Bending rigidity and Gaussian bending stiffness of single-layered graphene. *Nano letters* **2012,** *13* (1), 26-30.
51. Zou, X.; Liu, Y.; Yakobson, B. I., Predicting dislocations and grain boundaries in two-dimensional metal-disulfides from the first principles. *Nano letters* **2012,** *13* (1), 253-258.
52. Wang, H.; Li, Q.; Gao, Y.; Miao, F.; Zhou, X.-F.; Wan, X., Strain effects on borophene: ideal strength, negative Possion's ratio and phonon instability. *New Journal of Physics* **2016,** *18* (7), 073016.
53. Wu, J.; Wang, B.; Wei, Y.; Yang, R.; Dresselhaus, M., Mechanics and Mechanically Tunable Band Gap in Single-Layer Hexagonal Boron-Nitride. *Materials Research Letters* **2013,** *1* (4), 200-206.
54. Li, T., Ideal strength and phonon instability in single-layer $MoS_2$. *Physical Review B* **2012,** *85* (23), 235407.
55. Wei, Q.; Peng, X., Superior mechanical flexibility of phosphorene and few-layer black phosphorus. *Applied Physics Letters* **2014,** *104* (25), 251915.
56. Wang, Y.-J.; You, X.-R.; Chen, Q.; Feng, L.-Y.; Wang, K.; Ou, T.; Zhao, X.-Y.; Zhai, H.-J.; Li, S.-D., Chemical bonding and dynamic fluxionality of a $B_{15}^+$ cluster: a nanoscale double-axle tank tread. *Physical Chemistry Chemical Physics* **2016,** *18*, 15774-15782.

**Table I.** Calculated Young's modulus $C$, Poisson ratio $\sigma$, bending stiffness $D$ and mass density $\rho$ of borophenes and other typical 2D materials. $x$ and $y$ are two principal lattice directions. For a honeycomb lattice, x and y correspond to armchair and zigzag directions, respectively.

| Materials | $C_x$ (N/m) | $C_y$ (N/m) | $\sigma_x$ | $\sigma_y$ | $D_x$ (eV) | $D_y$ (eV) | $\rho$ (kg/m$^2$) |
|---|---|---|---|---|---|---|---|
| triangular | 399 | 163 | -0.23 | 0 | 4.76 | 1.39 | 7.73×10$^{-7}$ |
| $v_{1/12}$ | 208 | 161 | 0.09 | 0.08 | 1.33 | 0.92 | 6.78×10$^{-7}$ |
| α | 212 | 212 | 0.14 | 0.14 | 0.79 | 0.79 | 6.49×10$^{-7}$ |
| $v_{1/8}$ | 216 | 222 | 0.17 | 0.18 | 0.74 | 0.59 | 6.38×10$^{-7}$ |
| $v_{1/6}$ | 189 | 210 | 0.15 | 0.17 | 0.56 | 0.39 | 6.07×10$^{-7}$ |
| $v_{1/5}$ | 196 | 208 | 0.11 | 0.12 | 0.52 | 0.54 | 5.88×10$^{-7}$ |
| graphene | 342 | 342 | 0.17 | 0.17 | 1.46 | 1.46 | 7.55×10$^{-7}$ |
| h-BN | 273 | 273 | 0.21 | 0.22 | 0.97 | 0.97 | 7.53×10$^{-7}$ |
| MoS$_2$ | 125 | 125 | 0.25 | 0.27 | 9.14 | 9.14 | 3.04×10$^{-6}$ |
| silicene | 60 | 60 | 0.22 | 0.22 | 0.44 | 0.44 | 7.24×10$^{-7}$ |
| phosphorene | 90 | 24 | 0.16 | 0.72 | 5.21 | 1.35 | 1.35×10$^{-6}$ |

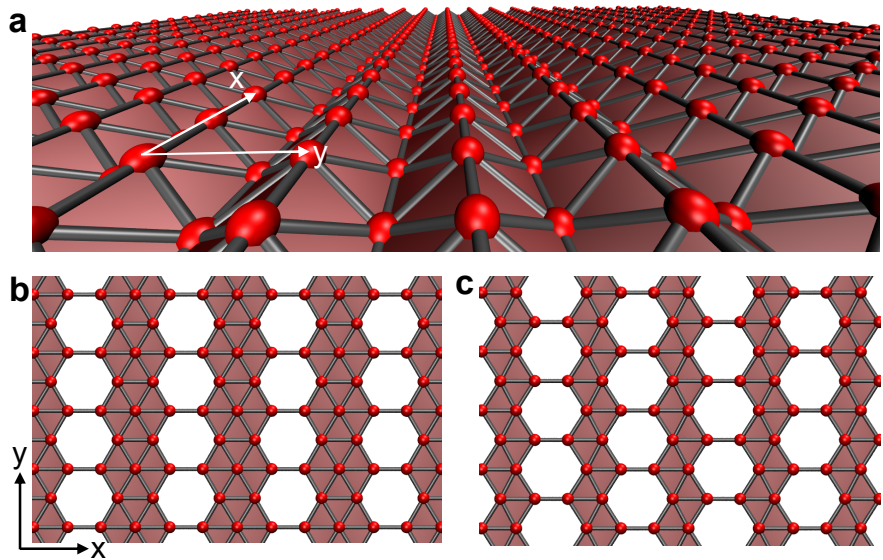

*Figure 1.* Atomic geometries of borophenes. (a) triangular, (b) $v_{1/6}$ and (c) $v_{1/5}$ sheets are shown.

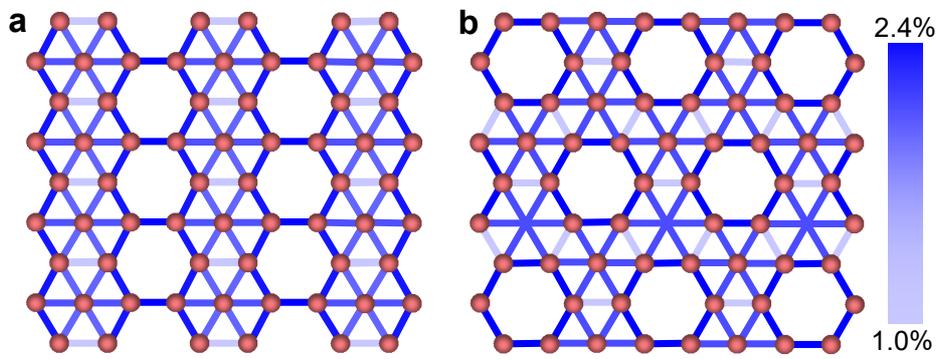

*Figure 2.* Maps of bond strains in the (a) $v_{1/6}$ and (b) α sheets under an applied biaxial strain of 2%.

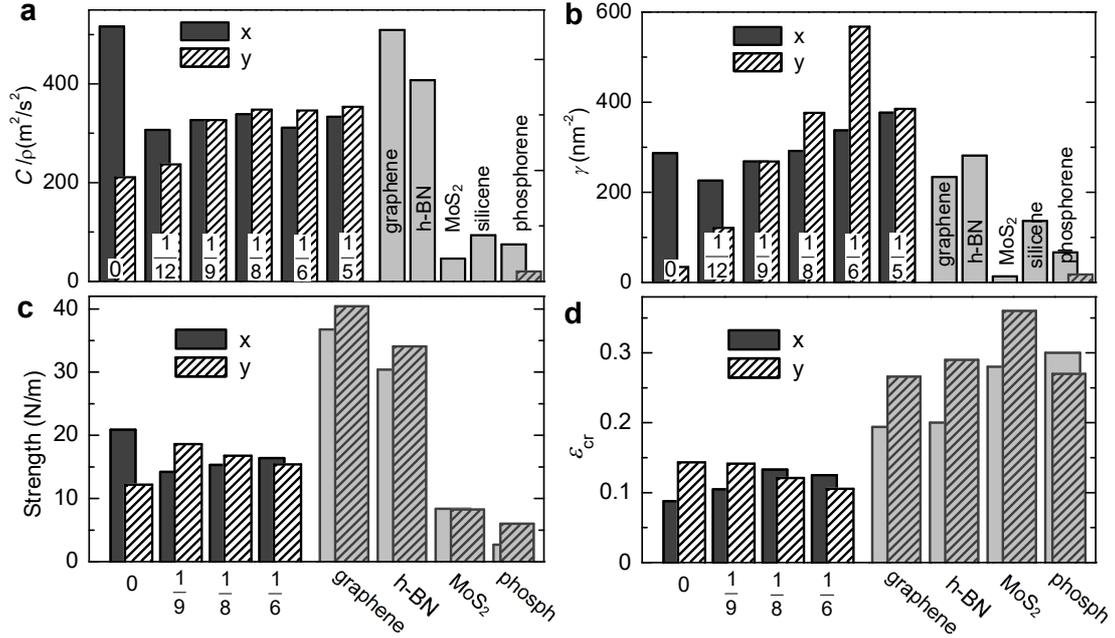

*Figure 3.* Elasticity and strength of borophenes. (a) Specific stiffness and (b) Foppl–von Karman number per unit area (i.e. the ratio between in-plane and bending modulus) of borophenes and other typical 2D materials. Collected ideal strength (c) and its critical strain (d) in borophenes, compared with values of other 2D materials. The data for h-BN sheet, $MoS_2$ and phosphorene in (c) and (d) are obtained from Refs. [53-55], respectively.

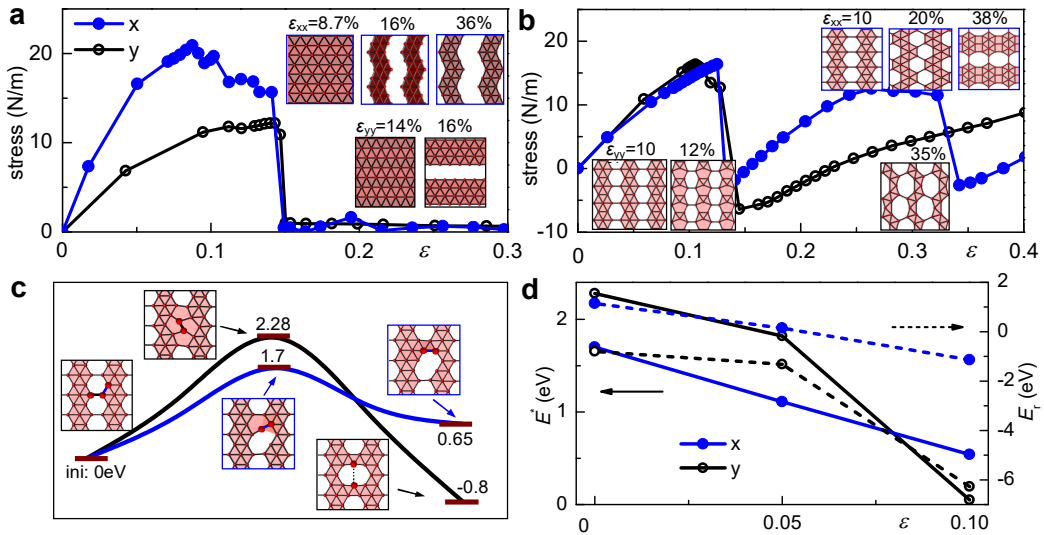

*Figure 4.* Stress-strain relationship. Tensile stress as a function of uniaxial strain in x and y directions for the (a) triangular and (b) $v_{1/6}$ sheets. (c) Minimum energy paths for two possible Stone-Wales bond rotations in the $v_{1/6}$ sheet (the rotated bonds are highlighted in thick black and blue, respectively). The inserts illustrate the atomic structures of initial, transition and final states. (d) Energy barrier and reaction energy for bond rotation as functions of applied strain. The blue line in (d) corresponds to the bond rotation shown in blue in (c).

13